\documentclass[preprint,aps]{revtex4}

\usepackage{graphicx}

\begin{document}

\title{Disappearance of Superconductivity and a Concomitant Lifshitz Transition in Heavily-Overdoped Bi$_2$Sr$_2$CuO$_{6}$  Superconductor Revealed by Angle-Resolved Photoemission Spectroscopy}
\author{Ying Ding$^{1,2}$,  Lin Zhao$^{1}$, Hong-Tao Yan$^{1,2}$, Qiang Gao$^{1,2}$, Jing Liu $^{1,2}$,Cheng Hu$^{1,2}$, Jian-Wei Huang$^{1,2}$, Cong Li$^{1,2}$, Yu Xu$^{1,2}$, Yong-Qing Cai$^{1,2}$, Hong-Tao Rong$^{1,2}$, Ding-Song Wu$^{1,2}$,  Chun-Yao Song$^{1,2}$,Hua-Xue Zhou$^{1}$, Xiao-Li Dong$^{1,2}$, Guo-Dong Liu$^{1}$, Qing-Yan Wang$^{1}$, Shen-Jin Zhang$^{3}$, Zhi-Min Wang$^{3}$, Feng-Feng Zhang $^{3}$, Feng Yang$^{3}$, Qin-Jun Peng$^{3}$, Zu-Yan Xu$^{3}$, Chuang-Tian Chen$^{3}$ and X. J. Zhou$^{1,2,4,5}$}

\affiliation{
\\$^{1}$Beijing National Laboratory for Condensed Matter Physics, Institute of Physics, Chinese Academy of Sciences, Beijing 100190, China.
\\$^{2}$University of Chinese Academy of Sciences, Beijing 100049, China.
\\$^{3}$Technical Institute of Physics and Chemistry, Chinese Academy of Sciences, Beijing 100190, China.
\\$^{4}$Songshan Lake Materials Laboratory, Dongguan, Guangdong 523808, China.
\\$^{5}$Collaborative Innovation Center of Quantum Matter, Beijing 100871, China.
}

\date{\today}

\begin{abstract}

By partially doping Pb to effectively suppress the superstructure in single-layered cuprate Bi$_2$Sr$_2$CuO$_{6+\delta}$ (Pb-Bi2201) and annealing them in vacuum or in high pressure oxygen atmosphere, a series of high quality Pb-Bi2201 single crystals are obtained with T$_c$ covering from 17 K to non-supercondcuting in the overdoped region. High resolution angle resolved photoemission spectroscopy (ARPES) measurements are carried out on these samples to investigate the evolution of the Fermi surface topology with doping in the normal state. Clear and complete Fermi surface are observed and quantitatively analyzed in  all these overdoped Pb-Bi2201 samples. A Lifshitz transition from hole-like Fermi surface to electron like Fermi surface with increasing doping is observed at a doping level of $\sim$0.35. This transition coincides with the change that the sample undergoes from superconducting to non-superconducting states. Our results reveal the emergence of an electron-like Fermi surface and the existence of a Lifshitz transition in heavily overdoped Bi2201 samples. They provide important information in understanding the connection between the disappearance of superconductivity and the Lifshitz transition in the overdoped region.
\end{abstract}


\maketitle
The normal state of high temperature cuprate superconductors, characterized by their unusual and distinct temperature dependence in the transport properties\cite{ATFiory1987MGurvitch,WFPeck1994HYWang,KKitazawa1995JMHarris}, is markedly different from the  usual metals that can be well described in terms of the Fermi liquid theory. The scattering process of the electrons around Fermi surface dominates the macroscopic physical properties of materials, therefore, the shape and size of Fermi surface play key roles in understanding the anomalous normal state properties of cuprate superconductors. In the underdoped, optimally-doped and even slightly overdoped regions, due to the existence of pseudogap\cite{BStatt1999TTimusk}, the quantitative determination of Fermi surface topology is complicated  from the debate between Fermi arc (gapless Fermi surface section) and Fermi pocket (small closed Fermi pocket)\cite{ZXShen1996DSMarshall,DGHinks1998MRNorman,ZXShen2005KMShen,JCCampuzano2006AKanigel,ZXShen2007WSLee,GDGu2008HBYang,LTaillefer2007NDLeyraud,NEHussey2008AFBangura,JRCooper2008EAYelland,XJZhou2009JMeng}.
Meanwhile,  the pseudogap is suppressed or absent in the heavily overdoped region\cite{JCCampuzano2011UChatterjee,JZaanen2015BKeimer} and it makes it possible to obtain complete Fermi surface in the samples that still exhibit unusual normal and  superconduting properties\cite{YShimakawa1992TManako,CTLin1996APMackenzie,XJZhou2010LZhao}. Assuming that the same superconductivity mechanism operates inside the superconducting dome, the study of the overdoped region provides an alternative and less complicated route in understanding the evolution between the superconducting state and non-superconducting state with doping.  In (La$_{2-x}$Sr$_x$)CuO$_4$ system, it was reported that its Fermi surface topology in the normal state changes from a hole-like pocket (centered at ($\pi$,$\pi$)) in the underdoped and optimally-doped regions to an electron-like pocket (centered at (0,0) in the overdoped region\cite{XJZhou_JESRP2002,SUchida2006TYoshida}. It is natural to ask whether this transition is universal or not, and whether there is a connection  between the Fermi surface topology change and the disappearance of superconductivity in the heavily overdoped region in other cuprate superconductors.  In the Bi$_2$Sr$_2$CaCu$_2$O$_8$ (Bi2212) system that has been  most extensively studied by angle resolved photoemission spectroscopy (ARPES), such a study has not been possible because it remains difficult to get Bi2212 samples with high enough doping that approaches the superconducting to non-superconducting transition. In comparison, the single-layer Bi$_2$Sr$_2$CuO$_6$ (Bi2201) system provides another desirable candidate for such a purpose.  It contains a single CuO$_2$ plane within one structural unit (half unit cell),  giving a single band and a single Fermi surface sheet that avoid band structure complications from multi-layered compounds where there is a band splitting\cite{ZXShen2001DLFeng,SIUchida2001YDChuang,SMaekawa2003SEBarnes}. In particular, it has been found that by partially substituting Bi with Pb in Bi$_2$Sr$_2$CuO$_{6+\delta}$ (Pb-Bi2201), together with annealing under different conditions, the samples can be pushed to the overdoped and heavily overdoped regions to become even  non-superconducting\cite{XJZhou2010LZhao}. Furthermore, the Pb-substitution can also suppress the incommensurate superstructure formation in Bi2201, thus  removing the electronic structure complications from the superstructure bands that occur in Bi2212 and other multi-layered bismuth systems. Therefore, Pb-Bi2201 provides an ideal system to investigate the electronic structure  evolution with doping in the  heavily overdoped region.

In this paper, a series of  Pb-Bi2201 single crystals have been prepared that covers the overdoped region (with a T$_c$  at 17 K) to heavily overdoped region, and to extremely overdoped region where the samples become non-superconducting. ARPES measurements are carried out to investigate the evolution of their Fermi surface topology with doping in the normal state. Clear and complete Fermi surfaces have been observed and quantitatively analyzed for all the samples. With the increase of doping accompanying the decrease of T$_c$, the Fermi surface of Pb-Bi2201 undergoes a Lifshitz transition changing from a hole-like pocket in the overdoped region to electron-like pocket in the extremely overdoped region. Our results reveal the emergence of electron-like Fermi surface in extremely overdoped Bi2201.  They also provide important information in understanding the connection between the disappearance of superconductivity and the Lifshitz transition in the overdoped region..

High quality  Bi$_{1.74}$Pb$_{0.38}$Sr$_{1.88}$CuO$_{6+\delta}$ (Pb-Bi2201) single crystals were grown by floating-zone technique\cite{XJZhou2010LZhao}. Plate-like single crystals with a size up to 5mm$\ast$10mm were obtained. The as-grown single crystals were post-annealed in different atmospheres, including in vacuum, flowing Ar, flowing air and high pressure oxygen at different temperatures (400$^\circ$C $\sim$ 600$^\circ$C) and for different times to change the doping level and to make the samples more homogeneous. All of the samples were quenched in liquid nitrogen right after the post-annealing in order to get sharp superconducting transition. The annealed samples were characterized by X-ray diffraction (XRD), magnetic susceptibility and electrical resistivity measurements, and the results for five representative samples with different superconducting transitions are shown in Fig.1. They are marked as OD17K for overdoped T$_c$ =17 K sample, OD11K for overdoped T$_c$ =11 K sample, OD7K for overdoped T$_c$=7 K sample, OD3K for overdoped T$_c$=3 K sample, ODNS for overdoped non-superconducting sample, respectively.  The OD11K and OD17K  samples were obtained by annealing the as grown Pb-Bi2201 sample in flowing Ar atmosphere and in vacuum, respectively, which can effectively remove  extra oxygen  to increase T$_c$.  The OD3K sample was obtained by annealing in air and the non-superconducting ODNS sample was obtained by annealing in high pressure oxygen atmosphere which can put more oxygen into sample to increase the hole doping.

High resolution angle-resolved photoemission measurements were performed on our photoemission system equipped with a Scienta R4000 electron-energy analyzer and a helium-discharge lamp, which gives a photon energy of h$\nu$ =21.218 eV\cite{XJZhou2008GDLiu}. The light on the sample is partially polarized. The energy resolution is set at 10$\sim$20 meV and the angular resolution is $\sim$ $0.3^\circ$ corresponding to 0.008 ${\AA}^{-1}$ momentum resolution at the photon energy of 21.218 eV.   All of the samples were cleaved \emph{in situ} at a low temperature of 30 K and measured in ultrahigh vacuum with a base pressure better than 5$\times$10$^{-11}$ mbar and at a temperature of 20 K in the normal state. The Fermi level is referenced by measuring on the Fermi edge of a clean polycrystalline gold that is electrically connected to the sample.

Figure 1(a) shows X-ray diffraction (XRD) patterns for some typical annealed Pb-Bi2201 single crystal samples which were measured by using a rotating anode x-ray diffractometer with Cu K$_{\alpha}$ radiation ($\lambda$=1.5418$\AA$). All the observed peaks can be indexed to the (00$l$) peaks of Bi2201, indicating a pure single-phase of the single crystals. The peaks are sharp, as exemplified from the (008) peaks in the top-left inset of Fig. 1(a) which have a width of $\sim$ 0.2$^\circ$ (full width at half maximum), indicating high crystallinity of the single crystals. The c axis lattice constant is shown in the top-right inset of Fig.1(a) which exhibits a monotonous decrease with T$_c$ but totally only 0.04\% change  is observed from the OD17K sample to the ODNS sample. This indicates the annealing process has a negligible effect on the lattice constant. Instead, it mainly changes the extra oxygen content in Pb-Bi2201. Figure 1(b) shows the temperature dependence of the magnetization for these Pb-Bi2201 single crystals measured under a magnetic field of 1 Oe. All the samples show narrow superconducting transition width within 1$\sim$2 K.  Figure 1(c) shows the temperature dependence of in-plane resistivity for the ODNS sample, measured using the standard four-probe method. The resistivity is found to be metallic with a concave shape that is commonly observed in heavily-overdoped samples such as La$_{1.7}$Sr$_{0.3}$CuO$_4$\cite{NEHussey2003SNakamae}.

Figure 2 shows the Fermi surface evolution of a series of Pb-Bi2201 samples with different dopings and different T$_c$s measured at a temperature of 20 K.  Figures 2(a)-2(e) show the Fermi surface mappings for five representative overdoped Pb-Bi2201 samples from OD17K, OD11K, OD7K, OD3K to ODNS samples, respectively. They are obtained by integrating the photoemission spectral weight within [-5, 5] meV energy window with respect to the Fermi level as a function of k$_x$ and k$_y$.  Fig. 2(f) shows the schematic Fermi surface that includes the main Fermi surface (black solid lines) and shadow Fermi surface (dashed lines) that are symmetrical with respect to the  (0,$\pm\pi$)-($\pm\pi$,0) lines in the Brillouin zone\cite{JFink2004AKoitzsch,KHirota2006Nakayama,MSGolden2006AMans}. All the samples show clear main Fermi surface (marked by red dashed lines in Figures 2(a)-2(e)) and shadow Fermi surface,and the spectral weight of the shadow Fermi surface is much weaker than that of  the main Fermi surface. All the observed features can be assigned to either the main Fermi surface or shadow Fermi surface with no sign of Fermi surface from superstructure modulation which are commonly observed in Bi2212. This further confirms that the incommensurate superstructure is well suppressed by Pb substitution in Pb-Bi2201\cite{XJZhou2010LZhao}.  For the OD17K sample (Fig.2(a)), its Fermi surface is hole-like with barrels centered around the ($\pi$,$\pi$) and its equivalent points. The typical feature is that two separated Fermi surface sheets cross around the (-$\pi$,0) antinodal region and its equivalent points nearly in parallel without any touching.  With the increase of doping and decreasing of T$_c$, the overall Fermi surface topology changes gradually and the major topology change occurs  around the antinodal region. From OD17K  to OD7K samples (Fig.2(b) and 2(c)), the two Fermi surface sheets near  (-$\pi$,0) get closer and closer, but their separation is still clear which indicates that they  still have  hole-like Fermi surface topology. For the OD3K sample (Fig.2(d)), these two Fermi surface sheets nearly touch at (-$\pi$,0) point.  For the ODNS sample (Fig.2(e)), it is clear that these two Fermi surface sheets merge together before approaching (-$\pi$,0) which indicates an electron-like Fermi surface topology  with barrels centered around the (0,0) point.  These results indicate that there is a critical doping between OD3K and ODNS samples, where a lifshitz transition\cite{Lifshitz1960} occurs with the Fermi surface topology changing from a hole-like to an electron-like topology.

To reveal the Fermi surface topology transition in more detail,  we focus our attention on the antinodal region around (-$\pi$,0). Figure 3 shows the Fermi surface and band structures around the antinodal regions for the OD17K and ODNS samples. For each sample, we carried out detailed momentum-dependent measurements around (-$\pi$,0) point (A1-A6 in Fig.3(a) for OD17K sample and B1-B6 in Fig.3(b) for ODNS sample). For the OD17K sample, small electron-like bands with the band bottom below the Fermi level can be observed for all the six momentum cuts (Fig.3(c)). Two-peak features are obvious in all their corresponding momentum distribution curves (MDCs) at the Fermi level in Fig.3(d). Especially for the key momentum cut A2,which crosses the (-$\pi$,0) point, both the measured band and the MDC at the Fermi level  clearly indicate the existence of two Fermi momenta k$_F$s.  These have provided  clear evidence that the OD17K sample has a hole-like Fermi surface topology. On the other hand,  for the ODNS sample, the situation is different. Small electron bands can only be observed for the bands of momentum Cut B4 to B6(Fig.3(e)). Their corresponding MDCs also exhibit two-peak feature in Fig.3(f). However, for the spectral images of Cut B3 to B1 (Fig.3(e)), which cover (-$\pi$,0) point, only a small patch is observed and the corresponding MDCs exhibit one peak feature. This provides clear evidence of an electron-like Fermi surface in the ODNS sample.

For the Bismuth-based cuprate superconductors, the band structures around the antinode region determine the Fermi surface topology. It is also essential to locate the Fermi momentum k$_F$ precisely at the nodal region  to get an accurate carrier concentration of the sample. Figure 4(a) shows the band structures along (-$\pi$,-$\pi$)-($\pi$,$\pi$) nodal direction crossing (0,0) point measured at 20 K for OD17K, OD11K and ODNS samples, respectively. These three images cover two main bands which should be centrosymmetric to the $\Gamma$(0,0) point. Their corresponding MDCs at the Fermi level are shown in Fig.4(b). The MDC peak positions determine the Fermi momentum k$_F$ along the nodal direction and the results are shown in the inset of Fig.4(b). Within our experimental uncertainty, the nodal k$_F$ exhibits only a slight change within our measured doping range.   Figures 4(c) and 4(d) summarize the photoemission images measured crossing the (-$\pi$,0) point and the corresponding MDCs at the Fermi level for all the five samples. With the increase of doping, the photoemission images evolve from an electron-like band to a small patch-like feature, and the corresponding two-peak MDCs change into a single peak feature.  This is consistent with the evolution of the Fermi surface topology from the hole-like to the electron-like. Figure 4(e) summarizes  all the Fermi surfaces quantitatively determined  in the first Brillouin zone. The corresponding carrier concentration (obtained according to the area of the Fermi surface) increases from $\sim$ 0.3 holes/Cu for OD17K to $\sim$ 0.4 holes/Cu (the inset in Fig.4(e)). Around the antinodal region, it is clear that there is a Lifshitz transition with increasing doping at $\sim$0.35 doping level.

It is well known that there is a close relationship between the Lifshitz transition and superconductivity, such as Lifshitz transition at the onset of superconductivity\cite{AKaminski2010CLiu},  collapse of the normal-state pseudogap at a Lifshitz transition\cite{AForget2015SBenhabib} and enhancement of superconductivity at a Lifshitz transition\cite{HDing2017XShi}. Our results indicate that the Lifshitz transition occurs at a critical doping somewhere between the OD3K and ODNS samples. For the OD3K sample at the antinode, the two Fermi surface sheets almost touch. For the ODNS sample, they already merge and the Fermi surface becomes electron-like.  It is interesting to note that  the Lifshitz transition observed in Pb-Bi2201 occurs at a critical doping where the superconductivity also disappears.  This observation may provide a possible connection between the disappearance of superconductivity and the Lifshitz transition in the Bi2201 system.

Our present results pose an interesting and challenging question on the origin of superconductivity in cuprate superconductors. Compared with another prototypical single-layer  La$_{2-x}$Sr$_x$CuO$_4$ (LSCO) system, Pb-Bi2201 is unique in two aspects.   First, for the LSCO, the Fermi surface change  from the hole-like to the electron-like occurs at a doping level of $\sim$0.16 holes/Cu which is only slightly overdoped\cite{XJZhou_JESRP2002,SUchida2006TYoshida}.  For the Pb-Bi2201 system, such a transition occurs at a much higher doping level $\sim$0.35. Second, when the Lifshtz transion occurs in LSCO, the sample is still superconducting. However, in Pb-Bi2201 system, the Lifshitz transition happens at the doping level where the sample becomes non-superconducting.  It is generally believed that superconductivity in cuprate superconductors is dictated mainly by doping the CuO$_2$ planes.  For the same single-layer systems, the doping dependence of superconductivity is very different between the LSCO and Pb-Bi2201 systems. The origin of such a dramatic difference is interesting and requires further investigation.

In summary, by carrying out high resolution angle-resolved photoemission spectroscopy measurements on Pb-Bi2201 to investigate the evolution of the Fermi surface topology with doping in the normal state, we have revealed a Lifshitz transition from a hole-like Fermi surface topology to an electron-like Fermi surface topology with increasing doping in the heavily overdoped Pb-Bi2201 system. The Lifshitz transition coincides with the disappearance of superconductivity. Our results provide a possible connection between the disappearance of superconductivity and Lifshitz transition in Bi2201 system. They also provide important information to understand the superconductivity in the overdoped region of high temperature cuprate superconductors.

\bibliographystyle{unsrt}

\vspace{3mm}

\noindent {\bf Acknowledgement}\\
This work is supported by the National Key Research and Development Program of China (Grant No. 2016YFA0300300) and 2017YFA0302900, the Strategic Priority Research Program (B) of the Chinese Academy of Sciences (Grant No. XDB07020300, XDB25000000), the National Basic Research Program of China (Grant No. 2015CB921300), the National Natural Science Foundation of China (Grant No. 11334010,11534007), and the Youth Innovation Promotion Association of CAS (Grant No. 2017013).

\vspace{3mm}

\noindent {\bf Author Contributions}\\
 X.J.Z., L.Z. and Y.D. proposed and designed the research. Y.D. and L.Z. contributed in sample growth and annealling. Y.D., L.Z., H.T.Y., Q.G.,  J.L.,C.H., J.W.H., Y.X., Y.Q.C.,H.T.R., D.S.W., C.Y.S.,G.D.L., Q.Y.W., S.J.Z., Z.M.W., F.F.Z., F.Y., Q,J,P., Z.Y.X., C.T.C. and X.J.Z. contributed to the development and maintenance of the Laser ARPES system and related software development. H.X.Z. and X.L.D. contributed to the magnetic measurement. Y.D.carried out the ARPES experiment with H.T.Y., Q.G., and L.Z.. Y.D., H.T.Y., L.Z. and X.J.Z. analyzed the data. Y.D., L.Z. and X.J.Z. wrote the paper. All authors participated in discussion and comment on the paper.\\

\vspace{3mm}


\newpage

\begin{figure}[tbp]
\begin{center}
\includegraphics[width=1.0\columnwidth,angle=0]{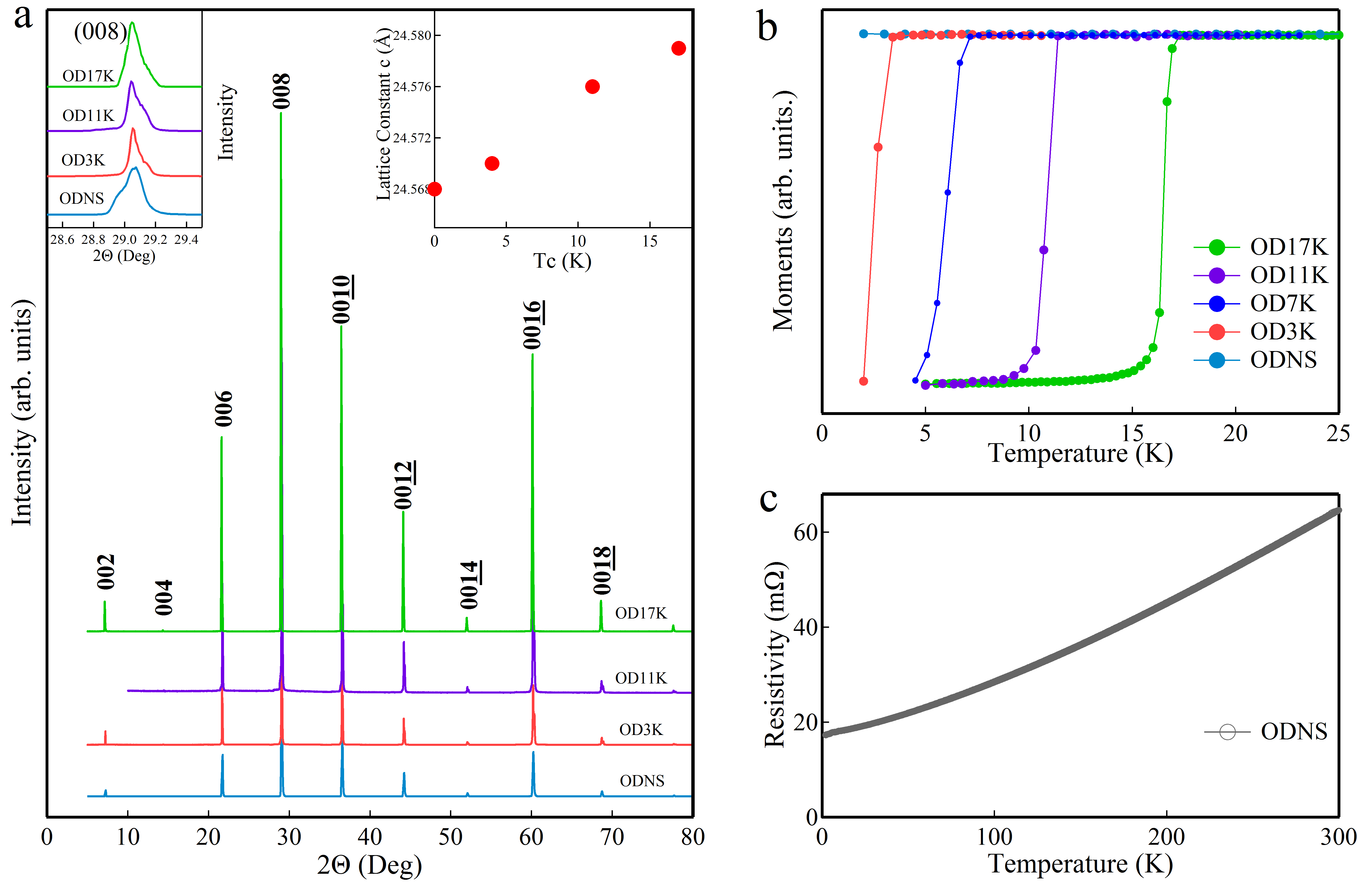}
\end{center}
\caption{Sample characterization of a series of annealed Pb-Bi2201 single crystals.
(a) X-ray diffraction  patterns for  Pb-Bi2201 single crystals with different T$_c$ obtained by annealing under different conditions.
The top-left inset shows the expanded (008) peak to highlight their position and the peak width. The c axis lattice constant is shown in the top-right inset. (b) Temperature dependence of the magnetization for the Pb-Bi2201 single crystals measured under a magnetic field of 1 Oe. (c) Temperature dependence of the in-plane resistivity for non-superconducting Pb-Bi2201 sample.
 }
\end{figure}

\begin{figure}[tbp]
\begin{center}
\includegraphics[width=1.0\columnwidth,angle=0]{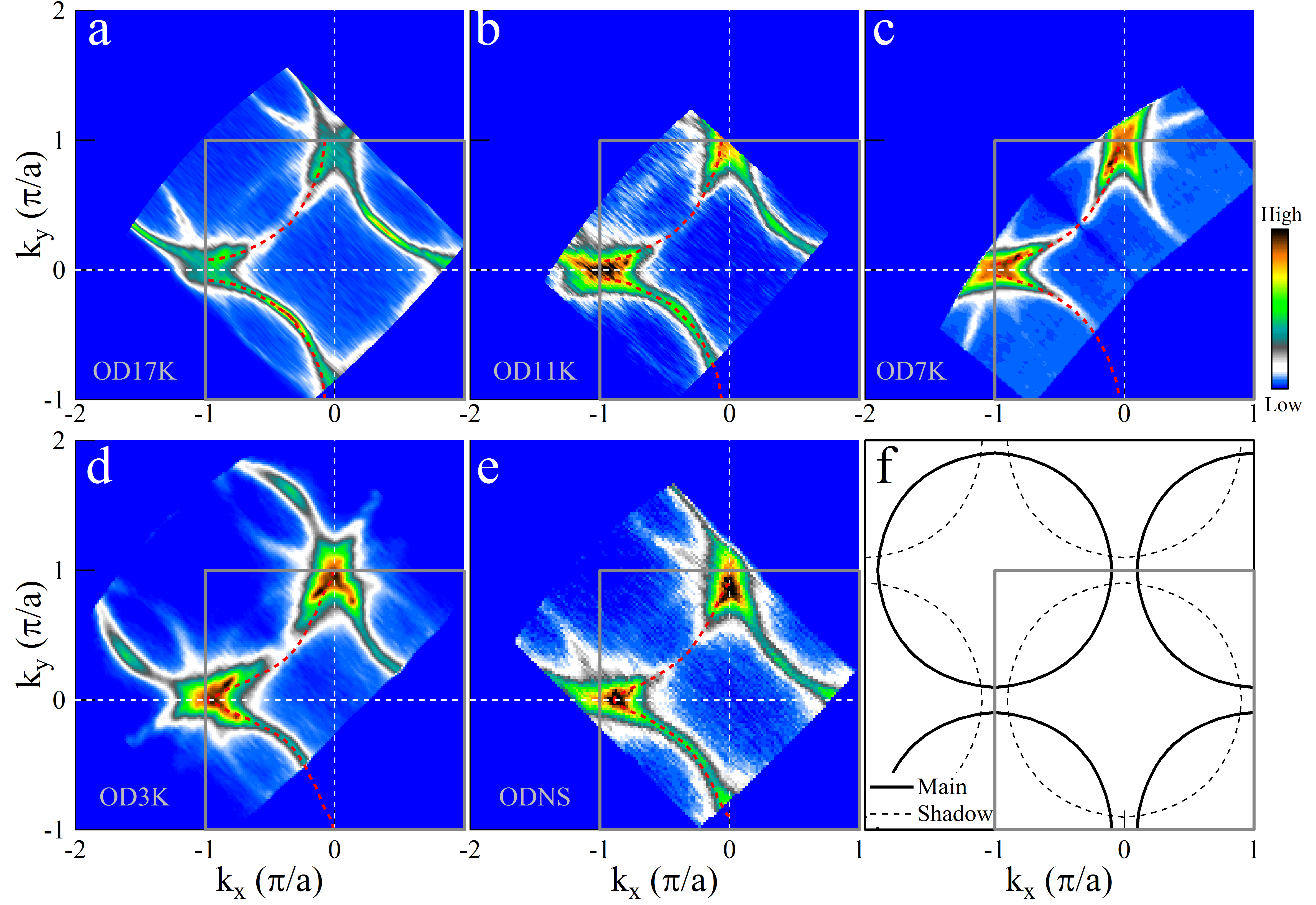}
\end{center}
\caption{Fermi surface evolution with doping of  Pb-Bi2201 samples.
(a)-(e) The Fermi surface for five representative overdoped Pb-Bi2201 samples (OD17K in (a), OD11K in (b), OD7K in (c), OD3K in (d) and ODNS in (e)), respectively,  by integrating within [-5, 5] meV energy window with respect to the Fermi level as a function of k$_x$ and k$_y$ measured at 20 K. The red dashed lines are guides to the Fermi surface topology. (f) shows the schematic Fermi surface of Bi2201 to understand the results in (a)-(e). Solid black lines represent the  Fermi surface from main bands and dashed lines represent the shadow Fermi surface which are symmetrical with main bands with respect to the  (0,$\pm\pi$)-($\pm\pi$,0) lines in the Brillouin zone.
 }
\end{figure}

\begin{figure}[tbp]
\begin{center}
\includegraphics[width=1.0\columnwidth,angle=0]{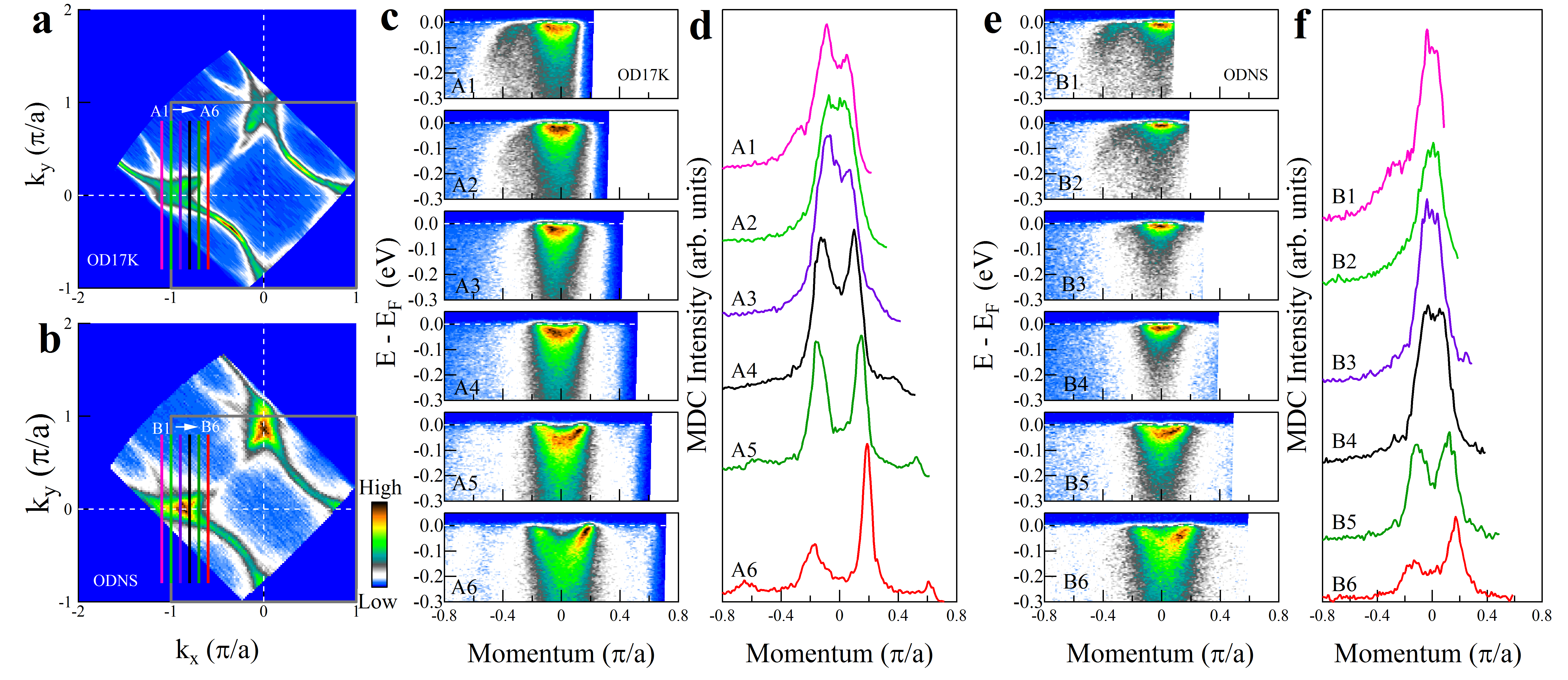}
\end{center}
\caption{Fermi surface and band structure for the OD17K and ODNS samples.
[(a),(b)] The Fermi surface of OD17K sample and ODNS sample measured at 20 K, respectively. (c) Band structure along six typical momentum cuts (cuts A1 to A6 as labelled in (a)) and the corresponding MDCs at the Fermi level are shown in (d) for the OD17K sample. [(e),(f)] Band structures along cuts B1 to B6 and the corresponding MDCs at Fermi level for the ODNS sample.}
\end{figure}

\begin{figure}[tbp]
\begin{center}
\includegraphics[width=1.0\columnwidth,angle=0]{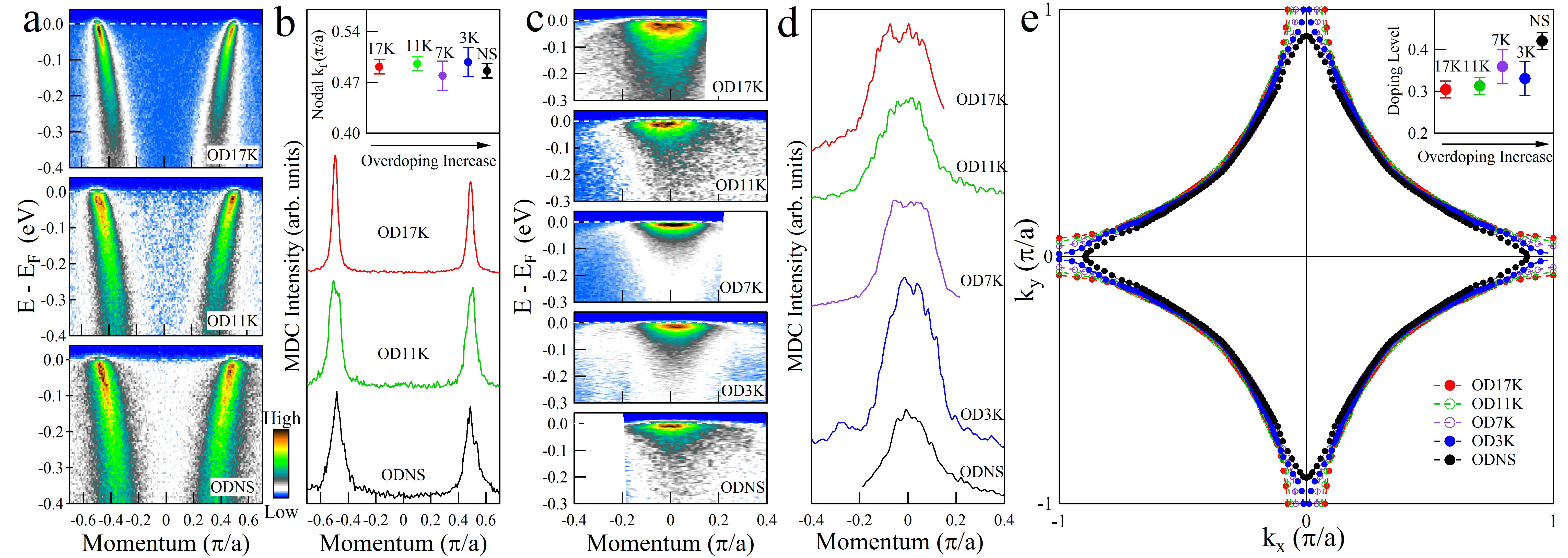}
\end{center}
\caption{Band structure along the nodal direction and antinodal direction for Pb-Bi2201 samples.
(a) The band structure along (-$\pi$,-$\pi$)-($\pi$,$\pi$) nodal direction crossing the (0,0) point measured at 20 K for  OD17K,  OD11K and ODNS samples, respectively. Their corresponding MDCs at the Fermi level are shown in (b). The Fermi momentum k$_F$s obtained by fitting MDCs are shown in the top inset of (b).   [(c),(d)] The band structure and  the corresponding MDCs at the Fermi level along the (-$\pi$,-$\pi$)-(-$\pi$,$\pi$) direction crossing (-$\pi$,0) point. All the Fermi surface are quantitatively obtained in the first Brillouin zone and summarized in (e). The corresponding carrier concentration for these five different samples is shown in the top-right inset of (e).
}
\end{figure}

\end{document}